\documentclass[12pt,preprint]{aastex}

\shorttitle{ 
The Boundary Layer of MV Lyr             
}
\shortauthors{Godon \& Sion}

\begin{document}

\title{
{\it FUSE} Observational Evidence of the Boundary Layer
of MV Lyrae in the High State 
} 

\author{
Patrick Godon \& Edward M. Sion 
} 
\affil{Astronomy \& Astrophysics, Villanova University, \\ 
800 Lancaster Avenue, Villanova, PA 19085, USA}
\email{
patrick.godon@villanova.edu   edward.sion@villanova.edu}

\begin{abstract} 

We carry out a spectral analysis of the archival {\it FUSE} spectrum  
of the VY Scl nova-like  cataclysmic variable 
MV Lyrae obtained in the high state.  
We find  that standard disk
models fail to fit the flux in the 
shorter wavelengths of {\it FUSE} ($\lambda < 950$\AA ).
An improved  fit is obtained by including a modeling 
of the boundary layer at the inner edge of the disk. 
The result of the modeling shows that in the high state  
the disk has a moderate accretion rate 
$\dot{M} \approx 2 \times 10^{-9} M_{\odot}$/yr, a low inclination, 
a boundary layer with a temperature of $\sim 100,000$K,  
and size $\sim 0.20R_{wd}$,
and the white dwarf is possibly heated up to a temperature $T \approx 50,000$K or 
higher. 

\end{abstract} 

\keywords{accretion, accretion disks - novae, cataclysmic variables
- white dwarfs}
 
\section{Introduction} 
 
\subsection{The VY Scl Nova-like System MV Lyrae} 

Cataclysmic variables (CVs) are evolved compact 
binaries in which the primary,
a white dwarf (WD), accretes matter and angular momentum from the
secondary, a star approximately on the main sequence 
filling its Roche-lobe. In dwarf nova
(DN) and non-magnetic nova-like (NL) systems (two classes of CVs), the matter
is transferred by means of an accretion disk around the          WD. 
DN systems are found mostly in a faint, quiescent state with a low  
mass transfer rate ($\dot{M}\sim 10^{-12}-10^{-11}M_{\odot}$/yr), and
every few weeks to months they transit into a bright, outburst state
lasting days to weeks,  
as the mass transfer suddenly increases by several orders of magnitude
($\dot{M}\sim 10^{-9}-10^{-8}M_{\odot}$/yr). The transition to outburst
in DNs is believed to be due to a disk instability \citep{sha86,can88,can98}.  
NL systems are found mainly in a high state phase,
characterized by a large mass accretion rate, and from time
to time the mass accretion rate suddenly drops by several
orders of magnitude. The transition from high state to low state
in NL is caused by a throttling of the mass transfer, possibly due
to magnetic activity of the secondary star (see \citet{war95} for
a review of CVs).  
With time, the accreted matter in CV systems forms a 
hydrogen-rich layer that covers the surface of the          WD. When enough
material is accreted (every few thousand years), 
the hydrogen-rich material undergoes a thermonuclear
runaway 
and the system forms a classical nova event.

MV Lyr is a VY Scl nova-like system, 
spending most of its time in 
the high state   
and occasionally undergoing short-duration drops in brightness. 
When this happens, 
the magnitude of MV Lyr drops from $V\approx 12-13$ to $V\approx 16-18$ 
\citep{hoa04}. From the AAVSO online data 
(http://www.aavso.org/) 
it appears that 
MV Lyr can stay in the high state for up to $\sim 5$yr, and then for a period
of a few years to $\sim10$yr it spends its time alternating between
high state and low state on a time scale of a few months to a year.  
The orbital period of the binary in MV Lyr is 
$P=3.19$hr and the mass ratio was found to be 
$q = M_{2nd}/M_{wd} =0.4$ \citep{sch81,ski95}. 
Because of the small
radial velocity amplitudes and the lack of eclipses, the
inclination is constrained to be in the range $i=10^{\circ}-13^{\circ}$,
though it has been remarked \citep{lin05} that 
the inclination could be as low as $i=7^{\circ}$. 

Observations of MV Lyr during a low state with the 
{\it International Ultraviolet Explorer ({\it IUE})} indicated a 
hot          WD possibly reaching 50,000K \citep{szk82} or higher \citep{chi82}. 
These observations were later
confirmed when 
MV Lyr was observed with {\it FUSE} on July 7, 2002, during a low state
that lasted for about 8 months. \citet{hoa04} carried out a spectral
analysis of the {\it FUSE} spectrum and found that the          WD must have 
a temperature of 47,000K, a gravity $\log{g}=8.25$, a projected rotational  
velocity $V_{rot} \sin{i} =200$km/s, subsolar abundances
$Z = 0.3 \times Z_{\odot}$, and a distance of $505 \pm 50$pc. 
At the time of the observations, the magnitude of MV Lyr was possibly as
low as $V\approx 18$ with a mass accretion rate no more than               
$\dot{M} \approx 3 \times 10^{-13}M_{\odot}$/yr \citep{hoa04}.  
A follow up analysis \citep{lin05} was carried out to study the 
different states of MV Lyr and its secondary, based  on       
archival {\it IUE} spectra, a HST/{\it STIS} snapshot 
and optical spectra. 
Using the parameters they derived in \citet{hoa04},   
\citet{lin05} found that standard disk models
did not provide adequate fits to the spectra; instead, the disk had to
be modified by, for example, a truncation of the inner disk, an isothermal
radial temperature profile in the outer disk, or both. 
These modified disk models provided some
satisfactory fit to the spectra obtained in the high state, 
leading to a mass accretion rate of the order of 
$3 \times 10^{-9}M_{\odot}$/yr. 

In the present analysis, the standard disk model fails again to provide 
a good fit to the observed ({\it{FUSE}}) spectrum of the system in the high state. 
This, however, points to a more general problem, namely that    
the standard disk model has failed to fit the 
observed spectra of many other CVs observed in the high state
\citep{pue07}. 
The direct  implication is that the standard disk model itself is inadequate in the ultraviolet (UV).  
It is in this context that we analyze here the {\it FUSE} spectrum of MV Lyr
in its high state. For this reason, we review next the
problems encountered in the modeling of CVs in the high state.  

\subsection{UV Spectral Modeling of CVs in the High State} 

In high state systems, the UV emission predominantly
originates from the disk,  and for that reason these systems became of special
interest, as they were expected to have an accretion disk
radial temperature profile given by the analytical expression 
$T_{eff}(r)$ of the standard disk model. More than two decades ago, 
\citet{wad88}  
used {\it IUE} spectra to show that CV NL systems 
(high state systems)  systematically disagree with the standard
disk model when either blackbody spectra or Kurucz stellar model
spectra were used to represent the accretion disk. A much improved 
version was developed by \citet{hub90} for modeling annuli of 
accretion disks that explicitly included calculation of
synthetic spectra using the standard disk model (the codes TLDISK/TLUSTY,
SYNSPEC and DISKSYN), and spectra of disks were computed 
\citep{wad98} assuming the standard disk model \citep{sha73,lyn74,pri81}.  
UV spectra (HST, HUT, {\it IUE, FUSE}) of DNs in outburst and
high brightness states of NLs were successfully modeled 
for a number of systems 
with synthetic spectra of optically thick disks 
\citep{kni02,ham07}.   

However, with time, it became clear that 
a significant number of systems still could not  be fitted  
using the standard disk model \citep{lon94,kni97,rob99} 
as first shown by \citet{wad88}.   
\citet{oro03} suggested a disk temperature profile ($T(r)$) flatter than the
standard disk model, to remediate at least partially to the problem,
invoking a possible non-steady mass accretion rate and/or irradiation
in the outer part of the disk. 
More recently, the UV spectra of RW Sex, V3885 Sgr, \& UX UMa   
\citep{lin08,lin09,lin10} 
were also modeled using a modified temperature profile, 
and it was suggested  that the outer disk temperature might
increase due to its interaction with the impacting L1 stream. 
Maybe the most significant of all is the 
statistical study of \citet{pue07} of 33 disk-systems which indicates that 
a revision of the standard model temperature profile, at 
least in the innermost part of the disk, is required. 

These studies concentrated  mainly on using          UV spectra 
(e.g. {\it IUE}, HST/{\it STIS}, {\it FUSE}), and the contribution to the
         UV continuum comes mainly from regions in the system where the
temperature ranges roughly between $\sim$10,000K and $\sim$100,000K. 
Lower temperatures peak in the optical, while higher temperatures peak
in the extreme UV. The          WD star and the disk are the main contributors to the
UV flux, while  the secondary star and the bright spot (the edge of the
disk impacted by the incoming stream of matter) peak in the optical 
and near UV, respectively. 
For disks in CVs with mass accretion rate $\sim 10^{-8}M_{\odot}$/yr, 
the inner disk reaches a temperature higher than 50,000K. 
These disk models, therefore, have a significant UV flux
contribution from the inner region of the disk.  

That same innermost part of the disk is the region 
where the matter in the disk has
to decelerate from its Keplerian motion 
in order to be accreted onto the more slowly rotating stellar surface,
as the centrifugal force keeps the matter from accreting.  
This region is known as the boundary layer (BL) and up to half
the accretion luminosity can be released in this final phase of 
the accretion process.  
Therefore, the emission and heating of the BL 
have to be taken into account when modeling an accretion disk
around a non-  (or weakly- ) magnetic star  
\citep{lyn74,pri81}.
It has now become clear that the temperature profile in the inner disk
is affected by the heating from the BL. Most importantly, BLs in high 
state have a temperature of the order of 100,000K-200,000K 
\citep{pop95,god95,pir04}, which is much lower than the first 
analytical estimates of $\sim$500,000K \citep{pri77,pri79}.   
Consequently, the BL of CV disk systems in the high state
contributes significantly to the UV, even though
it was predicted to emit only in the soft X-ray range 
(see also next section). 
Therefore, the BL has to be included in the disk modeling and the inner disk temperature
profile has to be modified accordingly.

It is the purpose of the present work to include an elementary (and preliminary)
model of the BL in the spectral modeling    
of the {\it FUSE} spectrum of MV Lyr taken during its high state.  
Since the BL
has an elevated temperature, its contribution will be more pronounced in 
the shorter wavelengths of the spectral range 
of {\it FUSE}.  

In the next section a brief review of the standard disk and   
BL models are given, and it is shown that the theory does indeed predict a 
BL 
contributing a non-negligible flux in the far          UV. 
The {\it FUSE} spectrum of MV Lyr together with
the details of the synthetic spectral modeling are presented in section 3. The results 
follow in section 4, and are discussed in the concluding section. 

\section{The Star-Disk Boundary Layer}

\subsection{The Standard Disk Model}

The total luminosity available through     
accretion is given by:  
\begin{equation} 
L_{acc} = \frac{G M_{*} {\dot{M}}} {R_{*}}, 
\end{equation}
where $G$ is the gravitational constant, $\dot{M}$ is the mass accretion
rate, $M_{*}$ is the mass of the
accreting star and $R_{*}$ is its radius. 
This expression simply reflects the release of
gravitational potential energy of the material brought onto the surface of
the star ($r=R_*$) from infinity ($r \rightarrow \infty$) at a rate
$\dot{M}$ (mass/time).

In the standard disk theory \citep{sha73,lyn74},
the accretion disk is axisymmetric, and geometrically thin
in the vertical dimension: it has a vertical thickness $H$ such that
$H/r << 1$. The disk is treated  with a one-dimensional radial approximation, 
in which the radial pressure is negligible and the matter in the disk is assumed to be
rotating at Keplerian speed.  
The 
energy dissipated by viscous processes between adjacent rings of matter
is instantly radiated locally in the vertical z-direction. 
The disk total luminosity $L_{disk}$ is half the accretion luminosity  
\begin{equation} 
L_{disk}=\frac{L_{acc}}{2} = \frac{G M_{*} {\dot{M}}} {2R_{*}},  
\end{equation} 
and each face of the disk radiates $L_{disk}/2$.   
This is so, because the matter at the inner edge of the disk ($r \approx R_*$) 
retains the remaining accretion energy in the form of 
kinetic energy, as it rotates at Keplerian speed.  

The effective surface temperature of such a disk is given by 
\citep{sha73,lyn74,pri81}:  
\begin{equation} 
T_{eff}(r) = T_0 x^{-3/4} ( 1 -x^{1/2})^{1/4}, 
\end{equation} 
where $x=r/R_*$, 
\begin{equation}  
\sigma T_0^4 = \frac{3 G M_* \dot{M}}{ 8 \pi R_*^3} , 
\end{equation}
and $\sigma$ the Stefan-Boltzmann constant.  
For practical purpose the last relation is usually written
\begin{equation} 
T_0 = 64,800K \times 
\left[ \left( \frac{M_*}{1 M_{\odot}} \right)  
\left( \frac{\dot{M}}{10^{-9} M_{\odot}/yr} \right)  
\left( \frac{R_*}{10^9 cm } \right)^{-3} \right]^{1/4} .
\end{equation}   
From this it is seen that accretion disks around          WD stars emit their  
energy in the optical and          UV. 
The above relation is obtained by assuming a no-shear boundary condition,
$\partial \Omega / \partial r =0$, at the stellar surface $r=R_*$
\citep{pri81}. 
Therefore, this model does not take into account the relatively 
slow rotation of the WD, and 
it  gives a disk temperature T=0K (!) at the stellar surface ($r=R_*$),
and a maximum temperature $T_{max}=0.488 T_0$ at $x=1.36$.
This model also neglects the spin up of the star by the disk.  

This relation, however, gives a good approximation of the
effective surface temperature in the disk  at larger radii
($r>>R_*$). For a typical high mass accretion rate 
of $\dot{M}= 1 \times 10^{-8}M_{\odot}$/yr, the maximum temperature in the
inner disk reaches about 50,000K-100,000K, and the inner disk become the strongest
emission source of the system in the UV. During the low state of NLs 
(or quiescent state of DNs) the mass  
mass accretion rate decreases down to $\dot{M} \approx 10^{-12} M_{\odot}$/yr,   
while the maximum temperature drops to less than $\approx$10,000K, making the
disk peak in the optical and barely emit any flux in the UV. 
The standard disk model is a good approximation for $r>>R_*$, but
in the inner disk region, one needs to model the BL between the
slowly rotating stellar surface and its Keplerian accretion disk.    
We would like also to note here that contrary to accreting neutron stars,
the irradiation of the disk by the heated WD and BL is 
negligible in accreting WDs \citep{van94,sha98,kin98}, and
one does not need to take it into account when modeling accretion disks
in CVs.  
 
\subsection{The Boundary Layer} 

Accreting WDs in CVs have stellar rotational velocities of the order of
a few 100km/s \citep{sio98} or about 10\% of the Keplerian velocity
at one stellar radius (a few 1000km/s).  
Consequently, the theory \citep{lyn74,pri81} predicts 
that the material at the inner edge of the disk rotating at Keplerian
speed has to adjust itself to the slowly
rotating stellar surface and, therefore,
dissipates its remaining rotational kinetic energy.
The remaining rotational kinetic energy liberated in the BL
($L_{BL}$) is nearly equal to half of the total
gravitational energy of the accreting matter ($L_{acc}/2=L_{disk}$), namely 
\citep{klu87}: 
\begin{equation}
L_{BL}= L_{disk}  \left(
1 - \frac{\Omega_{*}}{\Omega_K(R_{*})} \right)^2. 
\end{equation}
Because of its small radial extent and the large amount of energy
dissipated there, 
the BL was expected to emit only in the X-ray bands. 
At high mass accretion rates ($\dot{M} \approx  10^{-9}-10^{-8} M_{\odot}
yr^{-1}$) the BL should be optically
thick and its temperature was predicted to be in the range 200,000$K$ to 500,000$K$
\citep{pri77,pri79}. 
At low mass accretion rates, the BL should be
optically thin and emit in the hard X-ray (20keV), due to either
strong shocks or the formation of a X-ray 
emitting corona around the          WD \citep{pri79,kin84}. 

The first one-dimensional computations of the       BL       
\citep{reg83,pap86,sta87,reg88}  
were carried out by  assuming the ``slim disk equations''
\citep{szu88}: the equations are 
written and solved for vertically averaged quantities but also
include the transport of energy (radiation and advection) in the 
radial dimension.   
The results from these one-dimensional computations 
were in general agreement with the
predictions and showed that  
the BL should be narrow with a temperature increasing rapidly inward. 
However, more realistic 1D simulations 
\citep{pop95,god95,col97,col98a,col98b,col00a,col00b}  
showed that the BL extends radially and vertically more than  
expected. The result was that at high accretion rates the
BL is {\bf not} as hot as expected, with a temperature around 
125,000K \citep{god95} to 300,000K \citep{pop95}.                 
 
Numerical works in 2D \citep{rob86,kle87,kle89a,kle89b,kle91}  
did not clearly   reveal whether 
the accreting material accumulates in an equatorial belt 
(e.g. as claimed by \citet{dur77} and \citet{kip78})
or spreads quickly over the entire          WD surface  
\citep{mac83,liv87}. 
But more recently, using the semi-analytical 
shallow-water approximation developed by \citet{ino99} for
accreting neutron stars, \citet{pir04} found  
that at high mass accretion rate 
($\dot{M} > 10^{18}$g$~$s$^{-1} \approx 1.6 \times 10^{-8}M_{\odot}$/yr)
the combined effect of centrifugal force and pressure
gives rise to two latitudinal rings of enhanced brightness above
and below the           WD's equator: the ``spread layer'' 
\citep{ino99}.  
At lower mass accretion rate 
($\dot{M} < 10^{18}$g$~$s$^{-1} \approx 1.6 \times 10^{-8}M_{\odot}$/yr)
the spreading is negligible and most of the dissipated energy is radiated
back into the accretion disk. 
The effective temperature ($2 \times 10^4$K-$5 \times 10^5$K) and rotational velocity ($1 \times 10^8$cm/s-$3\times 10^8$cm/s) 
of the spread       BL       are similar to those of the one-dimensional       BL      
models (at least where they overlap in the parameter space). 
Some more recent simulations have been carried out to study 
the spread layer \citep{fis05,bal09}, however, these two-dimensional
simulations do not yet include radiation hydrodynamics. 

The significance of a cooler BL means that its emission  in the UV band 
is stronger and cannot be neglected when computing the luminosity of the
disk and          WD.  In the next subsection (sec.2.3) we summarize the 
      BL       emission one
may expect to observe in the UV at different mass accretion rates
based on the above mentioned investigations.

\subsection{The Emission of the Boundary Layer}

For $\dot{M} \approx 2 \times 10^{-8} M_{\odot}$/yr and higher, one expects 
the       BL       to spread onto the stellar surface, with a temperature
of the order of 
$2 \times 10^5$K-$5 \times 10^5$K, with a velocity of the order of $0.3-1 \times
V_K(R_*)$. The spread layer should be observable above and below the disk 
\citep{pir04}. At such temperatures the BL emits mainly in the soft X-ray
with some emission in the extreme UV, and possibly also far UV. The remaining part of
the WD not covered by the spread layer emits in the UV and is likely to have
an elevated temperature. The disk itself emits in the UV (far UV to near UV as the
radius increases) and optical (outer disk). As the mass accretion rate increases, 
the thickness of the boundary/spread layer increases and the matter becomes
advection-dominated as the heating rate surpasses the cooling rate. 
In this extreme  case,  
which is achieved when the mass accretion rate approaches its Eddington limit
\citep{god97}, 
the flow becomes spherically symmetric and the
slim disk approximation for a one-dimensional  BL becomes inaccurate. Instead,
the vertically averaged quantities represent  the spherically symmetric 
quantities \citep{nar95}. This implies that the results of the one-dimensional
advective       BL       in the plane of the disk represent a massive spread layer
engulfing the entire surface of the star. Such simulations have only been carried 
out for       BL       around FU Ori stars (e.g. in 1D \citet{pop93a,god96b}, and
in 2D \citet{kle96}) and symbiotics \citep{god96a}.

For an accretion rate in the range 
$3 \times 10^{-10} M_{\odot}$/yr  $ < \dot{M} <  2 \times 10^{-8} M_{\odot}$/yr,  
the matter does not spread far from the WD equator and most of the dissipated
energy radiates back into the accretion disk. The BL remains confined to the 
plane of the disk,  and one can expect
the one-dimensional simulations to accurately represent the BL    
between the star and disk. 
In this case, the exact temperature, size and rotation rate of the BL  
are given in \citet{pop95} and \citet{god95}. 
The temperature is in the
range $\approx 150,000$K \citep{god95} to a few 100,000K \citep{pop95}. 
A modeling of the extreme UV (Chandra) spectra of some CVs (e.g. OY Car, SS Cyg \&
WZ Sge) in this regime of $\dot{M}$ leads to such temperatures
\citep{mau00,mau04}.  
Here too,  the BL emits in the soft X-ray regime
with relatively more contribution 
to the extreme UV and far UV as the temperature decreases. The WD and the inner disk
emit in the UV, the outer disk emits in the optical. The modeling we
carry out in the present work is the modeling of such a  BL.

For  $\dot{M} <  3 \times 10^{-10} M_{\odot}$/yr, the BL is optically thin and 
the only UV emission is coming from the inner disk edge irradiated by the optically
thin BL. That inner edge forms a ring at a radius $r \sim 1.2R_* -1.6 R_*$,
rotating at Keplerian speed, with an effective temperature $T_{eff} \sim    
100,000$K$-50,000$K \citep{pop93b,pop99} 
for an accretion rate of $    3 \times 10^{-10} M_{\odot}$/yr
and   $    3 \times 10^{-11} M_{\odot}$/yr, respectively. As the mass accretion rate decreases, the
ring radius increases and its effective temperature decreases.  
The BL emits in the hard X-ray range and the disk's inner edge ring emits in the UV
if its temperature is high enough. Such UV emission from the inner
edge of the disk/BL was identified in VW Hyi \citep{pan03} and in other
dwarf novae in quiescence \citep{pan05}. This UV contribution from the
BL has to be taken into account when modeling the UV
spectra of accreting WDs in quiescence \citep{god05}.  
The WD emits in the UV, as does the very inner disk. The rest of the disk emits
in the optical. As the mass accretion rate decreases further, the disk eventually
ceases to emit in the UV.

As a consequence, the standard disk model is roughly correct for  
(say) $r \approx 2 R_*$ and larger, at both high and low mass accretion rates.  
The exact value of $r$ for which the standard disk model breaks down  
depends on the mass accretion rate and mass of the WD. 
At smaller radii the temperature in the disk-BL
region increases rapidly compared to the standard model prediction. 

\section{Data Analysis} 

\subsection{The {\it FUSE} Archival Data} 

MV Lyr was observed with {\it FUSE}
on May 6, 2003 (at 20h:07m:23s UT), just after it climbed into a high
state that lasted for about 4yr. 
The {\it FUSE} exposure (D9050901) consists of 7637s of good exposure time. 
The data were processed with the latest and final version of CalFUSE 
\citep{dix07} following  the same procedure as in previous works 
(for details see \citet{god09}).

The {\it FUSE} spectrum of MV Lyr (Figure 1) is characterized by broad
and deep absorption lines from highly ionized species and the absence 
of Hydrogen Lyman lines (except for sharp lines from the ISM).    
These absorption lines are 
N\,{\sc iv} ($\sim$923\AA\ ), 
S\,{\sc vi} (933.5\AA\ \& 944.5\AA\ ), 
O\,{\sc vi} (1131.6 \& 1137.3\AA\ ), 
S\,{\sc iv} (1062.6 \& 1073\AA\ ), 
Si\,{\sc iii} ($\sim$ 1108\AA\ ),  
P\,{\sc v} (1118\AA\ ), 
Si\,{\sc iv} (1122.5 \& 1128.3\AA\ ), and 
C\,{\sc iii} (1175\AA\ ).
Such deep and broad absorption lines are often observed in CVs in high
state, even in systems with a large inclination where the Keplerian velocity broadening 
is expected to produce a rather smooth continuum (e.g. see the {\it FUSE} spectrum
of V3885 Sgr in \citet{lin09}). 
These broad and deep absorption lines most likely 
form in a hot corona above the heated stellar surface, BL  
or disk.
Additional transition wavelengths have been marked on the spectrum in Figure 1 and    
can be associated with some spectral features; however, they are not clearly detected.  

There are a few sharp absorption lines from the interstellar medium such as 
Si\,{\sc ii} 1020\AA ,  C\,{\sc ii} 1036\AA , Ar\,{\sc i} 1048.2 \& 1066.7\AA , and    
Fe\,{\sc ii} 1145\AA .  
The sharp emission lines are artifacts due to air glow or direct reflection of
sunlight inside the telescope (e.g., such as the  C\,{\sc iii} 977\AA\ 
\& He\,{\sc ii} 1168\AA\ lines).

\subsection{TLUSTY and SYNSPEC Codes} 

The modeling of the {\it FUSE} spectrum is carried out in steps. 
First, spherically symmetric stellar atmosphere structure 
as well as the vertical structure of disk rings are computed
using the code TLUSTY  \citep{hub88}. 
For the stellar structure the input parameters are the surface gravity,
effective temperature and the surface composition of the star. 
For the composition it is often enough to take Hydrogen and 
Helium and neglect the metals as long their abundance is  
low (say similar to solar composition), in most cases this does
not affect the computed atmospheric structure. For high temperatures 
we use the NLTE option.   
For the disk rings the input is the local mass accretion rate,
mass of the accreting star, radius of the star, and radius of the 
ring and solar composition. 
Alternatively, one can input for each ring the effective
temperature, effective vertical gravity and the column mass at midplane.  
 
Second, the code SYNSPEC is used to 
derive the detailed radiation and flux distribution of 
continuum and lines \citep{hub95}.
SYNSPEC generates the output spectrum 
of the stellar atmosphere structure and disk vertical structure
computed by TLUSTY.  
The input for SYNSPEC is the same as for TLUSTY and it uses the 
structure computed by TLUSTY as an input. Composition
is specified here with an additional input file, and can include
species up to iron or even zinc. In the present modeling we 
included only H, He, C, N, O, Si, P, and S. 

In the next step the code ROTIN 
is used to account for rotational and instrumental 
broadening of the lines for the stellar synthetic spectrum, while the
code DISKSYN \citep{wad98} 
is used to combine the disk rings together (where we substitute  
the inner rings with hot BL rings if a BL is computed) and account
for rotational broadening due to Keplerian motion. Limb darkening is 
also included in the codes DISKSYN \& ROTIN. 

Boundary layer synthetic spectra are generated by computing inner disk rings with an
elevated temperature matching the BL temperature obtained
from theoretical estimates.   
The rings of the BL are then substituted to the inner disk rings. 
We then use a reduced $\chi^2_{\nu}$ fitting
technique to find the best fit to the observed spectrum.   
During the $\chi^2_{\nu}$ fitting process of the {\it FUSE}
spectrum taken in the high state the mass of the WD, 
the inclination of the system,  and the composition
are kept constant because they have been firmly established from
previous observations. 
The remaining parameters, i.e. the temperature of the WD, the mass
accretion rate, the number of BL rings included, 
and the temperature of the BL rings  
are allowed to vary as they are unknown. The distance to the system is obtained
{\it a posteriori} from the spectral fitting as an output parameter; 
resulting models with a distance that does not agreed with the assumed
distance are rejected (see next section). 

In the present work we use the following versions of the software:
TLUSTY202, SYNSPEC48, \& ROTIN4 for WD photospheric spectra runing
on a Linux platform (Cygwin-X) with a GNU FORTRAN 77 compiler; 
TLDISK195 (a previous variant of TLUSTY), SYNSPEC43, \& ROTIN3 and DISKSYN7
for disk and BL spectra runing on a Unix machine also using a GNU FORTRAN 77
compiler (the GNU FORTRAN 77 compiler runs slightly differently under Unix and
Linux operating systems). 
The software packages and user's guides 
are available online (free download) at http://nova.astro.umd.edu/ .

\section{Results}

In the following spectral modeling of the {\it FUSE} spectrum of MV Lyr in its
high state, as we derive the mass accretion rate and characterize the
BL of the system, we wish to use the mass of the WD and 
the distance to the system as given. These parameters were derived  
from the {\it FUSE} spectrum of the system in the low state  
by \citet{hoa04},  who obtained a WD surface temperature $T=47,000$K, 
a gravity $\log{g}=8.25$, non-solar composition $Z\approx 0.3 Z_{\odot}$, 
and a distance $d\approx 505 \pm 50$pc.   
Since these parameters are extremely important, we decided 
to confirm the results of \citet{hoa04} and we carried out 
a modeling of the {\it FUSE} spectrum of MV Lyr obtained in the low state. 
We obtained a WD surface temperature $T=44,000$K, a gravity $\log{g}=8.1$, 
a projected   stellar rotational velocity $V_{rot}\sin{i}=200$km/s,
non-solar composition $Z\approx 0.5 Z_{\odot}$, and a distance 
$d\approx 550 \pm 50$pc.   
The small difference in the parameters derived 
is due to the facts that (i) \citet{hoa04}
did not include rotational broadening; and (ii) \citet{hoa04} used
the mass-radius relation of \citet{ham61} for zero-temperature WDs
(giving $M_{wd}=0.73M_{\odot}$ for $\log{g}=8.25$),
while we used the C/O-core WD model of \citet{woo95} for a $\sim$45,000K WD.  
(giving $M_{wd}=0.73M_{\odot}$ for $\log{g}=8.1$ but $M_{wd} \sim 0.8M_{\odot}$
for $\log{g}=8.25$),
We note, however,  that our results are in complete agreement with the modeling 
carried out by E.M.S. {\it within} \citet{hoa04}'s work. Therefore, 
in the following we use the following parameters: $M_{wd}=0.8M_{\odot}$,
$d \sim 500-550$pc, $T_{wd} =44,000$K and above. The abundances
and stellar rotational velocity have no effects on the fitting to the
{\it FUSE} spectrum in the high state as it is dominated by emission
from the solar composition disk with a large Keplerian velocity broadening.   

In the fitting of the high state spectrum,  no attempt is made to model 
the broad and deep absorption lines. These lines form in a
hot corona above the disk/BL and cannot be modeled with the synthetic spectral 
code TLUSTY/SYNSPEC. The code models the lines created in the disk itself, 
but not in a hot corona above it. As a consequence, only the observed 
continuum is considered and the broad absorption lines are masked.

The          WD model used initially in the combined 
     BL        + disk +          WD fits to the
{\it FUSE} spectrum of MV Lyr in the high state has $T=44,000$K.
In the high state 
the          WD temperature could be elevated due to continuing accretion.  
Therefore, models with larger          WD temperature are also computed 
(50,000K and up) but they give very similar results. 
This is due to the fact that the WD is not the main 
emitting component of the UV spectrum. The details are given below.  
We  assume here $i=10^{\circ} \pm 3^{\circ}$ \citep{sch81,ski95,lin05}.

The first four models presented in Table 2 are  the 44,000K WD + standard disk 
models with increasing mass accretion rate. 
All these models do not have enough flux in the shorter wavelength range
to fit the {\it FUSE} spectrum.  The  model with  $\dot{M}=5  \times 10^{-9}M_{\odot}$/yr 
gives a distance of 666pc, too large to be acceptable, and with  
$\dot{M}=1 \times 10^{-8}M_{\odot}$/yr the distance becomes twice the 
value found in the low state. 
As an example, a model with $\dot{M} = 3.5 \times 10^{-9}M_{\odot}$/yr 
is shown in Figure 1,   giving a distance of 557pc.  
The flux deficiency in the shorter wavelengths is clearly seen.  The inclusion of 
a hotter          WD does not provide a significant improvement of the model fit.  

In order to increase the flux in the shorter wavelengths,  
the temperatures of the two inner rings of the standard disk model are  
modified to represent the      BL       .  The first 
ring is located at $r_1=1.05 R_*$ and the second is at $r_2=1.20R_*$. The
temperatures of the  rings ($T_1$ \& $T_2$ respectively) are  listed in Table 2. 
In the  standard disk model these temperatures are below 50,000K for the
accretion rate considered here. For the modeling of the      BL       , 
rings are computed with temperatures between 100,000K and 175,000K,  
in agreement with the BL models of \citet{god95} ($\sim 125,000$K) 
and \citet{pop95} ($\sim 180,000$K) for the mass accretion rates considered here.  
In doing so, one is able to construct two models of BL: thin (first ring only) 
and extended (first two rings). 

The standard disk model with a mass accretion rate of 
$3 \times 10^{-9}M_{\odot}$/yr fits the long wavelength end of the 
{\it FUSE} spectrum using the assumed distance to MV Lyr. For this reason 
models with $\dot{M} \ge 3 \times 10^{-9}M_{\odot}$/yr are not considered  
when including the BL, since the BL  increases 
the flux of the model and therefore its distance becomes too large. Also, since MV Lyr is in a high
state, mass accretion rates below $1 \times 10^{-9}M_{\odot}$/yr are not considered 
(these points are discussed in the section 5).   

First, an accretion disk with a  mass accretion rate of $1 \times 10^{-9}M_{\odot}$/yr
is considered, and the      BL        temperature and size are then varied. 
It is found that the best model fits lead to a distance
of only  350pc-400pc. Since the mass accretion rate is fixed to $1 \times 10^{-9}M_{\odot}$/yr, the
only way to increase the distance is to increase the contribution of the      BL       ,
by increasing its temperature and/or size.  
In doing so, the models that give an acceptable distance (of say at least $\sim 430$pc)
have actually too much flux in the shorter wavelengths, a sign that the BL contributes too
much flux.  Only a few  of these models are listed in Table 2.  
These models are not better than the standard disk models: their $\chi^2_{\nu}$ is as large
and/or their distance is too short. The best fit models are obtained for a thin (one ring)      BL       
with a temperature of $\sim$150,000K-175,000K. 
The distance for these models is, however, far
too short. The inclusion of a heated          WD does not improve the models and produces only a 
small increase in the distance.

Next, to obtain models with a larger distance, the mass accretion rate is increased  
to $2 \times 10^{-9}M_{\odot}$/yr. This has the effect of decreasing the relative flux contributed by  
the BL. These models agree with the assumed distance and
have a lower $\chi^2_{\nu}$, especially the two-ring BL models. 
The best one-ring model is for a BL temperature of 175,000K. This model has a relatively low
$\chi^2_{\nu}$ and agrees well with the distance. 
For the two-ring BL models, there is little difference 
(in terms of $\chi^2_{\nu}$) between a temperature of 100,000K, 125,000K and 150,000K 
(or any combination of those) as they produce $\chi^2_{\nu} \approx 1$ with $d\approx 500 \pm 50$pc.    
A 150,000K extended (2 rings) BL model with  
disk and          WD is shown in Figure 2. This model fits the flux
in the shorter wavelength range very well ($\lambda < 970$\AA\ , 
ignoring the broad absorptions of N\,{\sc iv} \& S\,{\sc vi}) but
produces too much flux around 980-1010\AA\ . Extended BL models with  a lower temperature (e.g. 100,000K) 
do not fit the shorter wavelengths as well, but on the other side they better fit the 970-1010\AA\ 
region and, consequently, they have the same $\chi^2_{\nu}$. 
In a last effort to improve the modeling, 
the temperature of the WD is increased to see how it affects the results. 
As the   WD temperature
increases the hotter BL models deteriorate slightly while 
the cooler BL models improve slightly.    
These models are listed in the lower part of Table 2
A fit with a 50,000K  WD, an extended 100,000K BL and a 
$2 \times 10^{-9}M_{\odot}$/yr accretion disk model is shown   in Figure 3.

\section{Discussion and Conclusions} 
 
Overall, the best fit to the {\it FUSE} spectrum of MV Lyr 
in its high state consists of a composite WD + disk + BL model      
with a mass accretion rate  $\dot{M} = 2 \times 10^{-9}M_{\odot}$/yr. 
There are two equally acceptable BL solutions, either a broad
extended (two-ring) BL with a rather low temperature (in the range 100,000K-150,000K), or a thin (one-ring) BL
with a higher temperature (175,000K). 
The WD temperature is probably higher than the 44,000K found in the low state, 
due to ongoing accretion but it affects only slightly  the quality of 
the fitting of the models, we discuss this further at the end of this 
section.    

Since the modeling of the {\it FUSE} spectrum generates more than one solution 
over a finite spectral range, one must consider
the bolometric luminosity of each component for comparison with equations (2) \& (6).  
For each model, the total luminosity of the BL is computed assuming it radiates
as a black body. This is a reasonable assumption, 
as in this regime the BL is most 
probably optically thick \citep{god95,pop95}. 
The two components of temperature $T_1$ \& $T_2$ (given in Table 2), 
and with radii $r_1=1.05R_*$ and $r_2=1.2R_*$ are added. One then simply uses the Stefan-Boltzmann law 
and area of each ring to find the total        BL      luminosity.   
The bolometric luminosity obtained is then compared with the theoretical expectation in eq.(6)
using eq.(2). For almost all the disk+BL models listed in Table 2 it is found that the computed bolometric
BL luminosity is larger than the disk luminosity. The discrepancy is even larger for the     
model with $\dot{M} = 1 \times 10^{-9}M_{\odot}$/yr. The only models for which the BL bolometric
luminosity is smaller than the disk luminosity (eq.6) is for an extended BL (two-ring model)
with a temperature as low as 100,000K. For that model the bolometric BL luminosity is 93\%
of the disk luminosity, implying a rotational velocity (eq.6) of less than 4\% its Keplerian value 
or about 135km/s.

With an inclination $i=10^{\circ}\pm 3^{\circ}$ and a projected rotational 
velocity of $200\pm 50$km/s (as derived in the low state), 
the (non-projected) rotational velocity should be in the range  
667km/s$< V_{rot}<$2050km/s, or $0.18 < \Omega_*/\Omega_K < 0.54$ 
(the lower limit is for a velocity of 150km/s and $i=13^{\circ}$, 
while the upper limit is for 250km/s and $i=7^{\circ}$).  This produces a
BL luminosity in the range $ 0.68 > L_{BL}/L_{disk} > 0.21$. 
For the BL model to agree with the upper limit, one could reduce the 
BL temperature (by 10,000K)  or increase the mass accretion rate 
(by 1/3, such a model is listed at the end of Table 2) or both. 
However, one notes that the error in the WD mass of $0.1M_{\odot}$
($\sim 0.2$ in $\log{g}$)  produces a computed $L_{BL}/L_{disk}$ between 
0.55 (for a $0.9M_{\odot}$ WD mass) and $\sim$1.5 (for a $0.7M_{\odot}$
WD mass).  Since the BL model used here is rather simplistic, 
a perfect agreement between the computed bolometric
luminosity of the BL and that computed from eq.6 is not expected. 
However, within the limits of the errors, the results of the BL model 
are consistent with a broad BL (of size $\sim 0.2 R_*$)  
with a rather low temperature ($\sim 100,000$K or slightly less) 
and a mass accretion rate of the order of $ \approx 2 \times 10^{-9} M_{\odot}$/yr
(or slightly larger). These models are located the lower part of Table 2 
and the fit improves (lower $\chi^2_{\nu}$ and better distance) as the temperature
of the WD increases, reaching a best fit for $T=70,000$K. Though the temperature
of the WD is certainly not that high, this certainly points to the fact that the
temperature of the WD is elevated during the high state reaching $\sim 50,000$K
or a little higher.

\section{acknowledgements} 
This work was supported by the National Aeronautics and Space Administration (NASA)
under grant number NNX08AJ39G issued through the Office of Astrophysics Data Analysis 
Program (ADP) to Villanova University. We have used some of the online data
from the AAVSO, and are thankful to the AAVSO and its members worldwide for making
this data public and their constant monitoring of cataclysmic variables.

\newpage

\begin{deluxetable}{ccl}
\tablewidth{0pt}
\tablecaption{System Parameters for MV Lyr}
\tablehead{
Parameter        & Value                & Reference      \\ 
}
\startdata
$M_{wd}$         & 0.73-0.8$M_{\odot}$ & this work; \citet{hoa04}                 \\ 
$R_{wd}$         & 7,440km             & based on models of \citet{woo95}   \\ 
$M_{2nd}$        & 0.3$M_{\odot}$      & \citet{hoa04}                \\ 
$i$              & $10^{\circ} \pm 3^{\circ}$  & \citet{sch81,ski95}    \\ 
                 &                     & \citet{lin05}    \\ 
E(B-V)           & 0                   & \citet{bru94}                 \\ 
 $P$             & 3.19hr              & \citet{ski95}          \\ 
 $d$             & $\sim 500-550$pc    & this work; \citet{hoa04}      \\ 
$\dot{M}$    & $2 \times 10^{-9}M_{\odot}$/yr & This work; \citet{lin05}  \\ 
\enddata
\end{deluxetable}

\clearpage

\begin{deluxetable}{ccccccccc}
\tablewidth{0pt}
\tablecaption{Accretion Disk \& Boundary Layer Synthetic Spectral Model Fits}
\tablehead{
Model & $\dot{M}\times 10^9$ &  $T_{wd}$ &  d     & $T_1$     &  $T_2$    & WD/disk & $\chi^2_{\nu}$ & Figure  \\ 
      & $(M_{\odot}/yr)$     & ($10^3$K) & $(pc)$ & ($10^3$K) & ($10^3$k) &   \%    &                &              
}
\startdata
disk    &  1.0               &   44      &  301   &  25.9     &   32.2    & 19/81     &  3.066       &       \\     
disk    &  2.0               &   44      &  428   &  30.7     &   38.3    & 10/90     &  1.923       &       \\ 
disk    &  3.5               &   44      &  557   &  35.4     &   44.1    & 6/94      &  1.379       &   1   \\     
disk    &  5.0               &   44      &  666   &  38.7     &   48.2    & 4/96      &  1.270       &       \\     
 ...    &  ...               &   ...     &  ...   &  ...      &   ...     & ...       &  ...         &   ...  \\ 
disk+BL d &  1.0               &   44      &  370   &  175      &   32.2    & 13/87     &  0.856       &       \\   
disk+BL 1 &  1.0               &   44      &  349   &  150      &   32.2    & 14/86     &  1.083       &       \\   
disk+BL 2 &  1.0               &   44      &  336   &  125      &   32.2    & 16/84     &  1.376       &       \\   
disk+BL 3 &  1.0               &   44      &  327   &  100      &   32.2    & 16/84     &  1.376       &       \\   
disk+BL 4 &  1.0               &   44      &  448   &  150      &  150      &  9/91     &  1.302       &       \\   
 ...    &  ...               &   ...     &  ...   &  ...      &   ...     & ...       &  ...         &   ...  \\ 
disk+BL d &  2.0               &   44      &  480   &  175      &   38.3    & 8/92      &  1.019       &       \\   
disk+BL 1 &  2.0               &   44      &  464   &  150      &   38.3    & 8/92      &  1.203       &       \\   
disk+BL 2 &  2.0               &   44      &  454   &  125      &   38.3    & 9/91      &  1.358       &       \\   
disk+BL 3 &  2.0               &   44      &  447   &  100      &   38.3    & 9/91      &  1.485       &       \\   
disk+BL 4 &  2.0               &   44      &  538   &  150      &   150     & 6/94      &  1.058       &  2    \\   
disk+BL 5 &  2.0               &   44      &  507   &  125      &   125     & 7/93      &  0.993       &       \\   
disk+BL 6 &  2.0               &   44      &  484   &  100      &   100     & 8/92      &  1.056       &       \\   
disk+BL da &  2.0              &   44      &  516   &  175      &   100     & 7/93      &  0.985       &       \\   
disk+BL 7 &  2.0               &   44      &  500   &  150      &   100     & 7/93      &  1.000       &       \\   
disk+BL 8 &  2.0               &   44      &  491   &  125      &   100     & 7/93      &  1.025       &       \\   
 ...      &  ...               &   ...     &  ...   &  ...      &   ...     & ...       &  ...         &   ...  \\ 
disk+BL da &  2.0              &   44      &  516   &  175      &   100     & 7/93      &  0.985       &       \\   
disk+BL da &  2.0              &   50      &  521   &  175      &   100     & 8/92      &  0.995       &       \\   
disk+BL 4 &  2.0               &   44      &  538   &  150      &   150     & 6/94      &  1.058       &  2    \\   
disk+BL 4 &  2.0               &   50      &  542   &  150      &   150     & 8/92      &  1.078       &       \\   
disk+BL 4 &  2.0               &   60      &  548   &  150      &   150     & 10/90     &  1.082       &       \\   
disk+BL 4 &  2.0               &   70      &  555   &  150      &   150     & 12/88     &  1.090       &       \\   
disk+BL 6 &  2.0               &   44      &  484   &  100      &   100     & 8/92      &  1.056       &       \\   
disk+BL 6 &  2.0               &   50      &  489   &  100      &   100     & 9/91      &  1.030       &   3   \\   
disk+BL 6 &  2.0               &   60      &  496   &  100      &   100     & 12/88     &  1.012       &       \\   
disk+BL 6 &  2.0               &   70      &  503   &  100      &   100     & 14/86     &  0.989       &       \\   
disk+BL 6 &  2.7               &   50      &  545   &  100      &   100     & 8/92      &  1.068       &       \\   
\enddata
\end{deluxetable}

\clearpage 

\begin{figure}[h] 
\vspace{-10.cm} 
\plotone{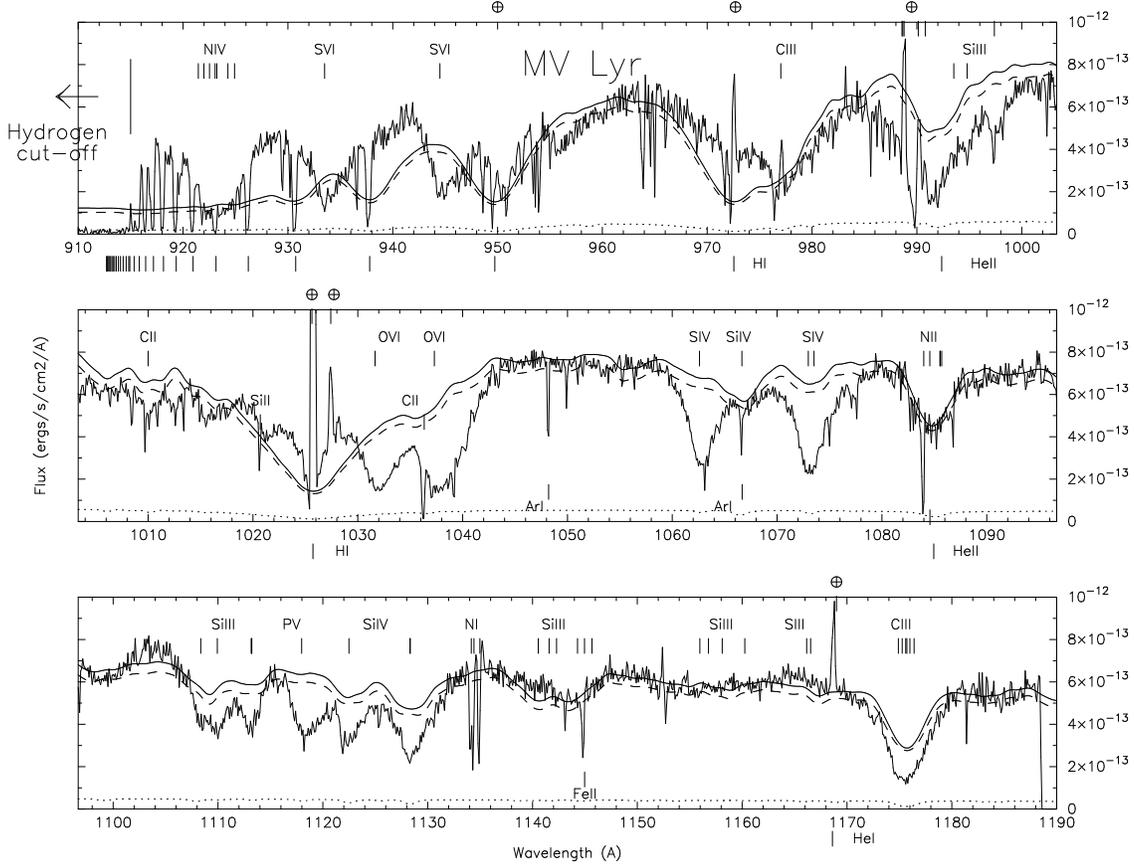}
\epsscale{1.2} 
\caption{The {\it FUSE} spectrum of MV Lyr in its high state  
is modeled with a 44,000K          WD (dotted line),
with a mass $0.8M_{\odot}$ and a standard accretion disk  (dashed line) 
with $\dot{M}=3.5 \times 10^{-9}M_{\odot}$/yr and $i=10^{\circ}$. 
The combined model          WD + disk is the solid black line.   
This gives a matching distance of 557pc.  
The model cannot match the flux level in the short wavelength range 
($<950$\AA ). In order to fit that region the mass accretion rate
has to be increased, and this increases the distance to twice
the distance of MV Lyr. 
} 
\end{figure} 

\clearpage

\begin{figure}[h] 
\vspace{-10.cm} 
\plotone{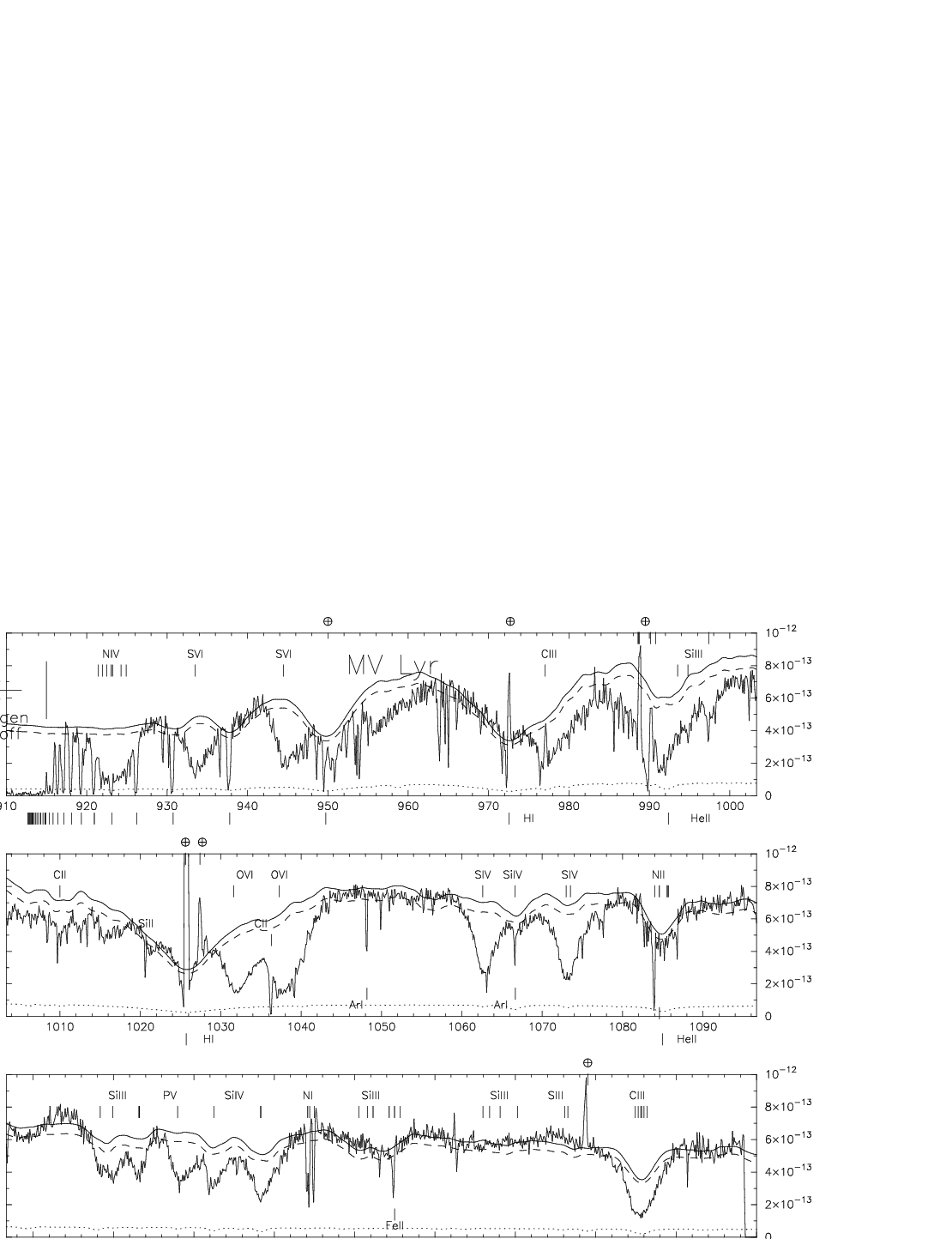}
\epsscale{1.2} 
\caption{The {\it FUSE} spectrum of MV Lyr in its high state  
is modeled with a composite          WD + disk  model which includes the        BL     . 
The          WD model (dotted line) has a
temperature of 44,000K and a mass $0.8M_{\odot}$, the accretion disk 
model (dashed line) has a mass accretion rate of 
$\dot{M}=2.0 \times 10^{-9}M_{\odot}$/yr and $i=10^{\circ}$. 
The        BL      is modeled as the two inner rings of the disk with  a temperature 
of 150,000K. 
The combined model          WD + disk is the solid black line.   
This gives a  distance of 538pc. The          WD contributes 
only 6\% of the flux in the {\it FUSE} spectral range.   
} 
\end{figure} 

\clearpage

\begin{figure}[h] 
\vspace{-10.cm} 
\plotone{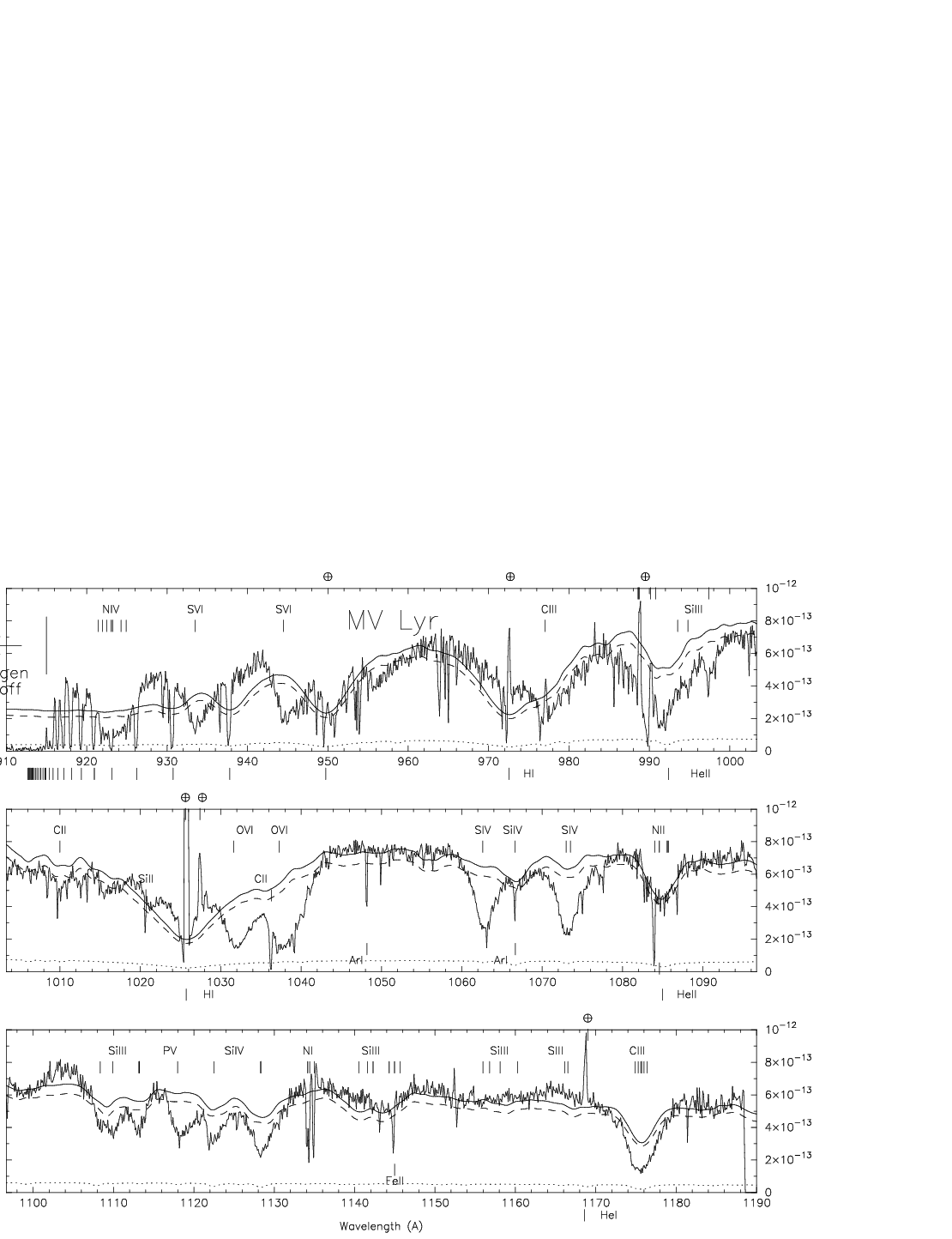}
\epsscale{1.2} 
\caption{The {\it FUSE} spectrum of MV Lyr in its high state  
is modeled with a composite          WD + disk  model which includes the        BL     . 
The          WD  model (dotted line) has a
temperature of 50,000K and a mass $0.8M_{\odot}$, the accretion disk 
model (dashed line) has a mass accretion rate of 
$\dot{M}=2.0 \times 10^{-9}M_{\odot}$/yr and $i=10^{\circ}$. 
The        BL      is modeled as the two inner rings of the disk with  a temperature 
of 100,000K. 
The combined model          WD + disk is the solid black line.   
This gives a  distance of 489pc. The          WD  contributes 
9\% of the flux in the {\it FUSE} spectral range.   
} 
\end{figure}

\end{document}